\documentclass{iau} 
\usepackage{graphicx}

\title[S321.~~Mid-infrared outskirts of M31] 
{Stellar populations in the outskirts of M31: the mid-infrared view}

\author[P. Barmby \& M. Rafiei Ravandi]   
{P. Barmby$^1$, M. Rafiei Ravandi$^1$}

\affiliation{$^1$Department of Physics and Astronomy and Centre for Planetary and Space Exploration,
University of Western Ontario, London, Canada \\ email: {\tt pbarmby@uwo.ca}} 

\pubyear{2016}
\volume{321}  
\jname{Formation and Evolution of Galaxy Outskirts}
\editors{A. Gil de Paz, J. C. Lee \& J. H. Knapen, eds.}
\begin{document}

\maketitle

\begin{abstract}
The mid-infrared provides a unique view of galaxy stellar populations, sensitive to both the integrated light of old, 
low-mass stars and to individual dusty mass-losing stars. We present results from an extended Spitzer/IRAC survey 
of M31 with total lengths of 6.6 and 4.4 degrees along the major and minor axes, respectively. The integrated surface 
brightness profile proves to be surprisingly difficult to trace in the outskirts of the galaxy, but we can also investigate 
the disk/halo transition via a star count profile, with careful correction for foreground and background contamination. 
Our point-source catalog allows us to report on mid-infrared properties of individual objects in the outskirts of M31, 
via cross-correlation with PAndAS, WISE, and other catalogs.

\end{abstract}

\firstsection 
\section{Introduction}

Getting the full picture of the stellar populations in the nearest large galaxy, M31, poses challenges because of its
large angular extent on the sky. The PAndAS project \cite[(McConnachie \etal\ 2009)]{pandas} covers
a huge area around M31 in visible light, but observations at other wavelengths are needed to tell the whole story.
As seen by the S$^4$G survey, the mid-infrared has the advantages of tracing the integrated light of low-mass stars while being relatively unaffected 
by interstellar extinction \cite[(Querejeta \etal\ 2015)]{q15}. Mid-infrared point-source observations can 
be used to identify  mass-losing stars with circumstellar dust \cite[(Boyer \etal\ 2015)]{boyer15} and red
supergiants \cite[(Britavskiy \etal\ 2015)]{brit15}. Here we present preliminary results on the stellar populations
observed in a wide-field mid-infrared survey of M31.

\section{Observations}

\begin{figure}[t]
\begin{center}
 \includegraphics[width=5in]{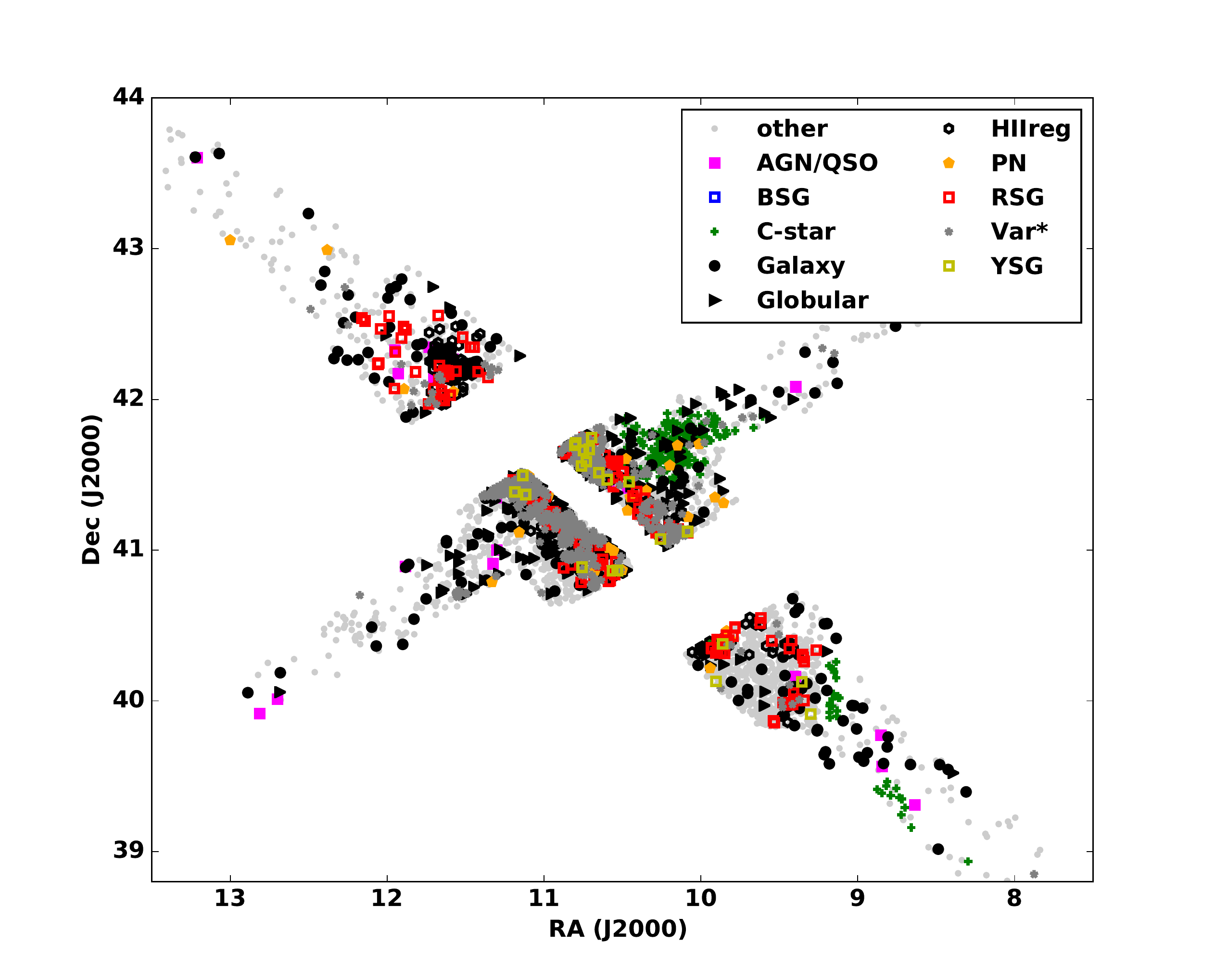} 
 \caption{Sky positions of point sources in the outskirts of M31 matched to SIMBAD sources with a tolerance of 1~arcsec. North is up and East to the left.
 Blank regions along the major (NE-SW) and minor (NW-SE) axes are locations where crowding was too severe for accurate point-source extraction.
The concentration of galaxies in the northeast part of the disc is due to identification in a study of M31 star clusters by the PHAT project 
\cite[(Johnson \etal\ 2012)]{johnson12} 
while the carbon stars to the northwest were found in a study of the satellite galaxy NGC~205 by \cite[Demers, Battinelli \& Letarte (2003)]{dbl03}.}
   \label{fig1}
\end{center}
\end{figure}

With about 40 hours of {\it Spitzer} observations in the post-cryogenic mission, we extended the IRAC 3.6 and 4.5~$\mu$m
mapping of M31 presented by \cite[Barmby \etal\ (2006)]{barmby06} to larger distances along the major and minor
axes of the galaxy (Fig.\,\ref{fig1}). \cite[Rafiei Ravandi \etal\ (2016)]{rr16} describes the observations, the challenging process of background
subtraction, and construction of a point-source catalog containing over $4\times 10^5$ individual sources. The catalog reaches (Vega)
magnitude of $\approx 19$ at both 3.6 and 4.5~$\mu$m, about 1 magnitude fainter than the tip of the red giant branch (TRGB) luminosity
at the distance of M31. Using wide-field, shallow surveys to model Galactic foreground and narrow-field, deep surveys to model 
unresolved background galaxies, we developed a scheme to estimate the probability that 
a given mid-infrared source belonged to M31 from its magnitude and colour. Probable M31 AGB stars
are found  at $[4.5]\lesssim15$ and $[3.6]-[4.5] \geq 0.4$.

\section{Analysis: mid-infrared point sources in M31} 

Our extensive point-source catalog allowed for detailed comparisons between the stellar populations as seen in the IRAC
3.6 and 4.5~$\mu$m bands and those observed at other wavelengths. Comparing with 2MASS and WISE  \cite[(Rafiei Ravandi \etal\ 2016, Fig.\ 6)]{rr16}, 
we find that sources detected only by IRAC and WISE detections are redder and fainter than those detected by 
all three of IRAC, WISE, and 2MASS. This is consistent with the greater depth and redder response of WISE compared to 2MASS.
Comparing with PAndAS \cite[(Rafiei Ravandi \etal\ 2016, Fig.\ 7)]{rr16}, we find that IRAC+PAndAS detections are optically-redder and more likely to be
Galactic foreground  dwarfs compared to the PAndAS-only detections; however, the IRAC observations do detect some probable M31 red giants
as expected from the detection limit below the TRGB.

Both the SIMBAD and NED databases include individual components of nearby galaxies (e.g., stars, star clusters, planetary nebulae, H{\sc ii} regions),
although these are not the main focus of either database. We matched the IRAC sources against SIMBAD objects with a
distance tolerance of 1~arcsec, with the results shown in Figures~1 and 2.
Figure~1 shows that, as expected, the sources
categorized as stars or star-related (PNe, H{\sc ii} regions) are strongly concentrated toward the disk of the galaxy, while background
galaxies and active galactic nuclei are more uniformly distributed. Globular clusters fall in between these two distributions and
are easily detectible in this dataset.

\begin{figure}[t]
\begin{center}
 \includegraphics[width=4.5in]{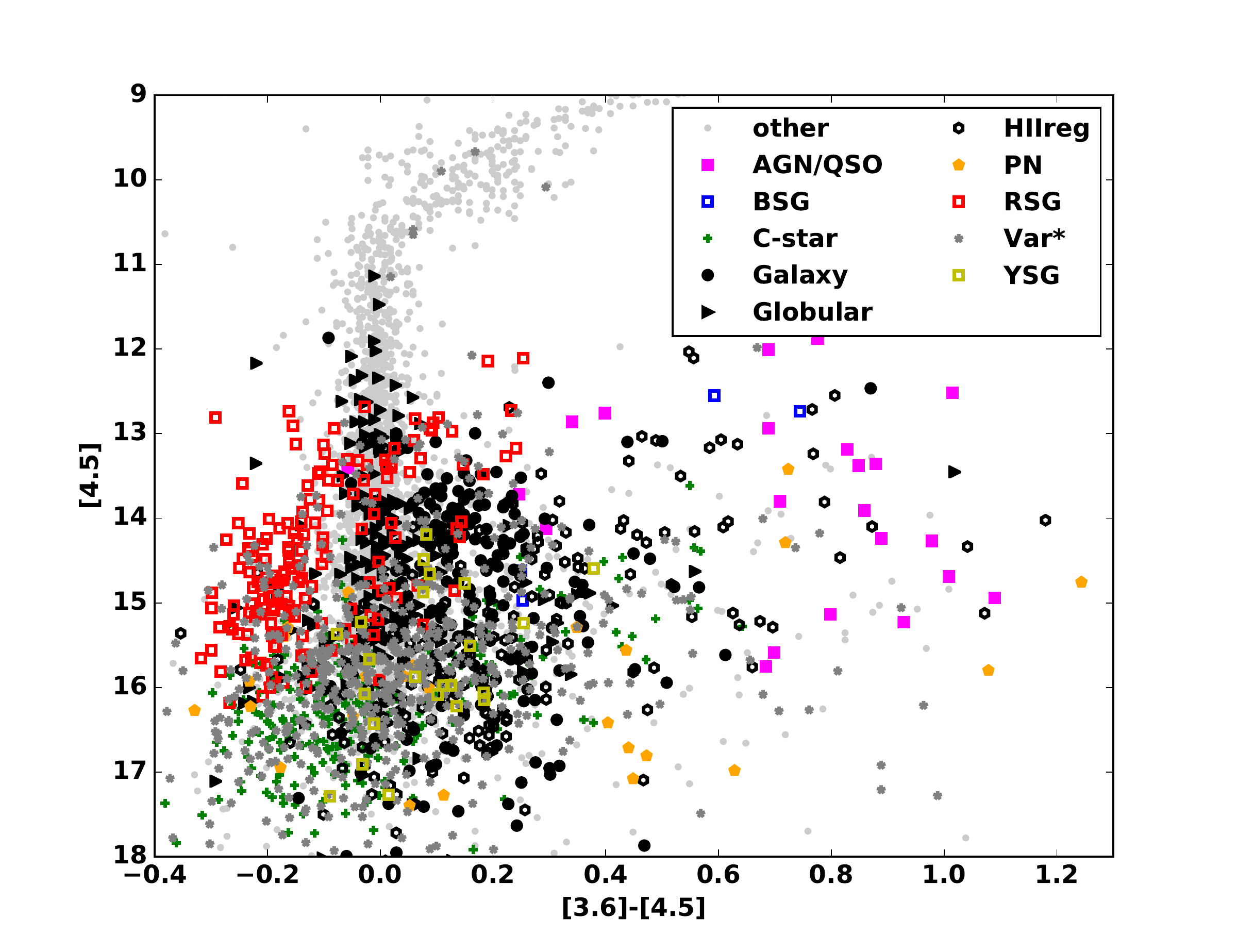} 
 \caption{Spitzer/IRAC colour-magnitude diagram showing point sources in the outskirts of M31 matched to SIMBAD
 sources with a tolerance of 1~arcsec. All magnitudes are in the Vega system; the red colours of objects brighter than
 $[4.5]\approx 11$ are due to saturation in the 3.6~$\mu$m band. Some of the SIMBAD sub-classes (e.g. of variable stars)
have been combined.}
   \label{fig2}
\end{center}
\end{figure}

Figure~2 shows the colour-magnitude diagram of matched sources. As expected from deep-field extragalactic surveys,
the active galactic nuclei and quasars have very red $[3.6]-[4.5]$ colours; background galaxies are mostly found at $[3.6]-[4.5]>0.1$.
Globular clusters have Vega-magnitude colours  $[3.6]-[4.5]\approx 0$; see also \cite[Barmby \& Jalilian (2012)]{bj12}.
Individual stars in M31 show a wide range of mid-infrared colours; interestingly, (optically) red supergiants are mostly found at $[3.6]-[4.5]<0$
while the handful of blue supergiants have $[3.6]-[4.5]>0.2$, so `red' and `blue' are reversed in the mid-infrared.
\cite[Britavskiy \etal\ (2015)]{brit15} found similar results for supergiants in Local Group dwarf irregulars.
Combining mid-infrared surveys with those from other wavelengths should provide a rich variety of targets for spectroscopic follow-up
with the James Webb Space Telescope.

\end{document}